\documentstyle[12pt,psfig]{article}
\def\reference{\parskip 0pt\par\noindent\hangindent 0.5 truecm}

\begin{document}
%
%
\title{AAT/WFI observations of the Extragalactic H I Cloud HIPASS J1712-64}
%


\author{Geraint F. Lewis$^{1}$, 
Michael J. Irwin$^{2}$,
Rodrigo A. Ibata$^{3}$ \&
Brad K. Gibson $^{4}$
} 

\date{}
\maketitle

{\center
$^1$ Anglo-Australian Observatory, P.O. Box 296, Epping, NSW 1710, Australia\\
gfl@aaoepp.aao.gov.au\\[3mm]
$^2$ Institute of Astronomy, Madingley Rd, Cambridge, CB3 0HA, UK\\
mike@ast.cam.ac.uk\\[3mm]
$^3$ Observatoire de Strabourg, 11, rue de l'Universite, F-67000, Strasbourg, 
France \\ ibata@pleiades.u-strasbg.fr\\[3mm]
$^4$ Centre for Astrophysics \& Supercomputing, Swinburne University, 
PO Box 218, Hawthorn, Victoria 3122,
Australia\\bgibson@astro.swin.edu.au\\[3mm]
}

%
\begin{abstract}
AAT/WFI optical  images of a candidate extragalactic  HI cloud, HIPASS
J1712-64, are  presented.  The g  and r-band CCD mosaic  camera frames
were processed  using a  new data pipeline  recently installed  at the
AAO.  The  resultant stacked images reach  significantly deeper levels
than those  of previous published  optical imaging of  this candidate,
providing a  detection limit  $M_g\sim-7$ at a  distance of  3Mpc, the
inferred distance  to HIPASS J1712-64.  However,  detailed analysis of
the images fails to uncover any stellar population associated with the
HI emission. If this system is a  member of the Local Group then it is
pathologically  different to  other members.  Hence,  our observations
reinforce  earlier  suggestions  that  this  HI cloud  is  most  likely
Galactic in origin and not a Local Volume dwarf galaxy.
\end{abstract}

{\bf Keywords: galaxies: dwarf, Local Group -- techniques -- image processing}

\bigskip

%
%

\newcommand{\hvc}{J1712-64}

\section{Introduction}
All-sky  HI surveys have  found significant  numbers of  High Velocity
Clouds  (HVCs).   Although  various  scenarios have  been  invoked  to
explain  the  disparate  properties  of  the  ensemble  of  HVCs,  one
population  in  particular,  the  isolated  compact  HVCs,  remains  a
possible candidate for the continuing rain of material accreting on to
the  Milky Way  and other  Local Group  galaxies (e.g.Braun  \& Burton
1999;  Blitz et  al.  1999;  Sembach  et~al.  2002),  as predicted  by
models of hierarchical structure formation (e.g. Klypin et al. 1999).
 
The goal  of the current market  leader, the HI  Parkes All-Sky Survey
(HIPASS),  is to  map out  the  distribution of  HI gas  in the  local
universe, providing an  accurate map of the HVC  distribution at finer
resolution and to better sensitivity levels than obtained previously.

\begin{figure}
\begin{center}
\psfig{file=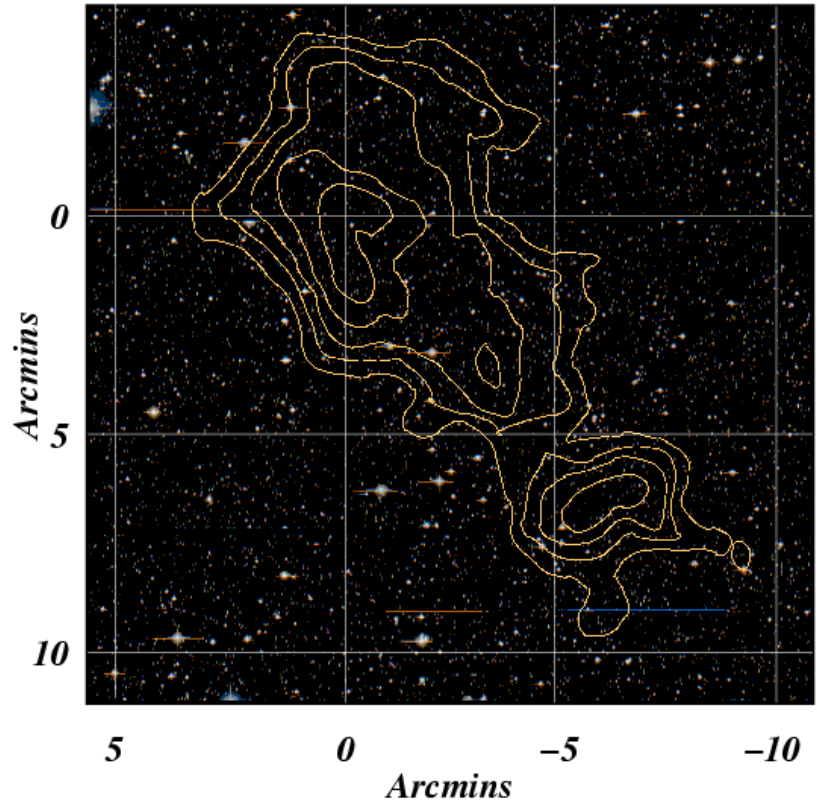,height=12cm,angle=-90}
\caption{ A  false colour image of  the HVC region  obtained using the
pipeline-processed  stacked $g$  and  $r$ band  images  to generate  a
colour optical image  of the region.  The HI  contours from Kilborn et
al.   (2000) are overlaid.   There is  no obvious  optical counterpart
visible in  either resolved stars down  to $r = 24,  g = 25$,  or as a
diffuse low  surface brightness  enhancement down to  a level of  $r =
28$,  $g =  29$ magnitudes  arcsec$^{-2}$ respectively.   The  grid is
relative to the centre of the CCD  [17 12 35.7, -64 38 12 (J2000)] and
the area covered in this figure is about $15 \times 15$ arcmin.
\label{Figure1}}            
\end{center}
\end{figure}

Recently, Kilborn et al. (2000) announced the detection of an isolated
compact  HI   cloud,  HIPASS  \hvc,   with  a  systemic   velocity  of
$\sim$450\,km\,s$^{-1}$.  Transforming  to the reference  frame of the
Local  Group,  \hvc's   velocity  is  $\sim$240\,km\,s$^{-1}$;  which,
assuming  the cloud  is  extragalactic,  places it  at  a distance  of
$\sim$3Mpc  (H$_\circ$=75\,km\,s$^{-1}$\,Mpc$^{-1}$).  Kilborn  et  al
(2000) provided  further evidence  for this  conclusion  using optical
imaging of \hvc,  with the goal of detecting  a stellar component.  To
limits  of  $R\sim19.2$ and  $B\sim19.8$,  no  overdensity of  stellar
sources were detected in the vicinity of \hvc.  However, at a distance
of 3\,Mpc, the brightest putative asymptotic giant branch (AGB) or red
giant  branch stars  (RGB) typically  found in  nearby  dwarf galaxies
would be  at $R$  magnitudes of $\sim$23--24  well below the  limit of
their optical  data.  Any unresolved low  surface brightness signature
due to  fainter stars  would also be  close to their  limiting surface
brightness  for detection,  making  the null  optical result  somewhat
ambiguous.

In an effort to acquire significantly deeper optical imaging we imaged
the region around \hvc\ with the Wide Field Imager (WFI) mosaic camera
at  the  AAT.  This  paper  presents the  new,  deep  images of  \hvc\
obtained  from these  observations.  The  observations  were processed
using a new data-pipeline at the Anglo-Australian Observatory, a brief
synopsis of which is given in Section~\ref{datareduction}. An analysis
of  the images  is  presented in  Section~\ref{imageanalysis}, and  we
finish  with  a  brief  discussion  and  summary  of  the  results  in
Section~\ref{discussion}.

\section{Data Reduction}\label{datareduction}
The data  were obtained on the  nights of the 17-19  August 2001 using
the  Wide   Field  Imager,   mounted  at  the   prime  focus   of  the
Anglo-Australian  Telescope at Coonabarabran  (see Tinney  2001).  The
1/2 degree $\times$ 1/2 degree field of view of WFI on the AAT make it
an ideal  instrument for searching  for optical counterparts  of HVCs.
For the  observations of HIPASS  \hvc\ the seeing was  sub-arcsec, but
high  intermittent   cirrus  made  accurate   photometric  calibration
difficult, and  the calibration  of the colour-magnitude  diagrams and
various  associated limits  alluded to  later  in the  paper are  only
accurate at the $\sim$0.1 magnitude level.

A total  integration time  of 2700\,s for  both $g$ and  $r$-bands was
acquired, split as 9$\times$300\,s  exposures in each band.  The field
was centred  at 17 12  35.7, -64 38  12 (J2000), corresponding  to the
centre of the  northern component of \hvc\ (Kilborn  et a. 2001).  All
the data were processed through  a prototype AAO/WFI pipeline based on
the data pipeline developed for  the Isaac Newton Telescope Wide Field
Camera  (INT  WFC).  The  pipeline  provides  the  standard tools  for
removing  the instrumental  signature of  CCD images  and additionally
provides means  for defringing the data (if  needed), object catalogue
generation  and the  ability  to provide  astrometric and  photometric
calibration in an  automated manner (see Irwin \&  Lewis 2001 for more
details).

As  an  example of  the  versatility of  the  pipeline,  the main  end
products in this case were a  pair of stacked $g$ and $r$-band images,
based on an  accurate World Coordinate System (WCS)  to coordinate the
stacking,   and   the    generation   of   a   deep,   detected-object
colour-magnitude diagram.

\section{Image Analysis}\label{imageanalysis}
A  false colour  image  of the  CCD  encompassing \hvc,  with the  HI
contours overlaid, is shown in figure~\ref{Figure1}, using r, 0.5(g+r)
and g  for the  RGB colours.  Careful  visual inspection of  the whole
mosaic  image  (four times  the  area  shown in  figure~\ref{Figure1})
reveals  no compelling  evidence  for any  dwarf  galaxy, resolved  or
otherwise, that could be associated with the HI envelope.

\begin{figure}
\centerline{\psfig{file=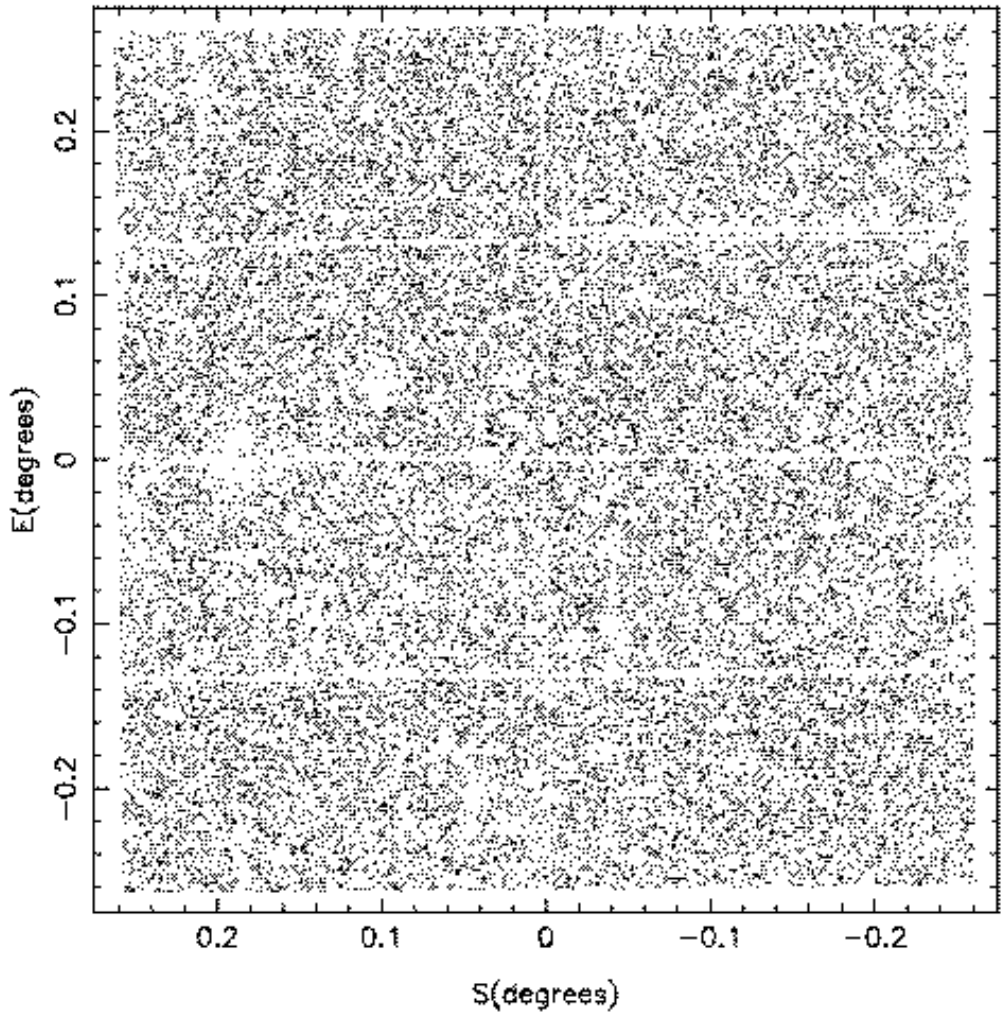,height=15cm,angle=-90}}
\caption{ The spatial distribution of stars in the range $22 < r < 24$
and $0.5 <  g-r < 2.0$ for  the WFI image of the  HVC region.  Visible
artifacts  include the small  gaps between  the 8  CCDs making  up the
mosaic  imager,   regions  affected   by  bad  columns,   and  regions
effectively masked  out by the light pollution  from bright foreground
stars in  the field.  The  overall distribution of stars  is extremely
uniform,  with no evidence  for any  underlying resolved  dwarf galaxy
population.
\label{Figure2}}            
\end{figure}

Morphological Classification as part of the data-reduction pipeline is
based  on  automatic  analysis   of  a  discreetly  sampled  intensity
curve-of-growth of  all images.  In  any pairwise combination  of such
measures, stellar images lie on a well-defined (and constant) locus as
a function of  magnitude leaving only the spread  (sigma) in the locus
to be further derived.   This greatly facilitates image classification
based on the  sigma-normalised distance from the locus.   The size and
"sign" of the  deviation(s) are used to assign images  to one of three
morphological classes: stellar, non-stellar, and noise-like.

At  Galactic coordinates,  $l  = 326.6^\circ,  b  = -14.6^\circ$,  the
extinction  in this  direction  is  relatively low  at  E(B-V) =  0.11
(Schlegel et  al. 1998), hence  all the standard  Galactic populations
should  be readily  visible, as  well as  large numbers  of background
galaxies and  any possible nearby dwarf  galaxy component.  Therefore,
we can be more quantitative in our search by using the distribution of
detected stellar images in $g$,$r$  to probe for any non-uniformity in
image distribution. Analysis of  isopleth maps, however, revealed that
all are  consistent with noise and  therefore there is  no evidence of
any resolved compact stellar population  in \hvc\ as would be expected
for   a    nearby   dwarf   galaxy.    An   example    is   shown   in
figure~\ref{Figure2} for the range $22<r<24$ and $0.5<g-r<2.0$.

\begin{figure}
\begin{center}
\psfig{file=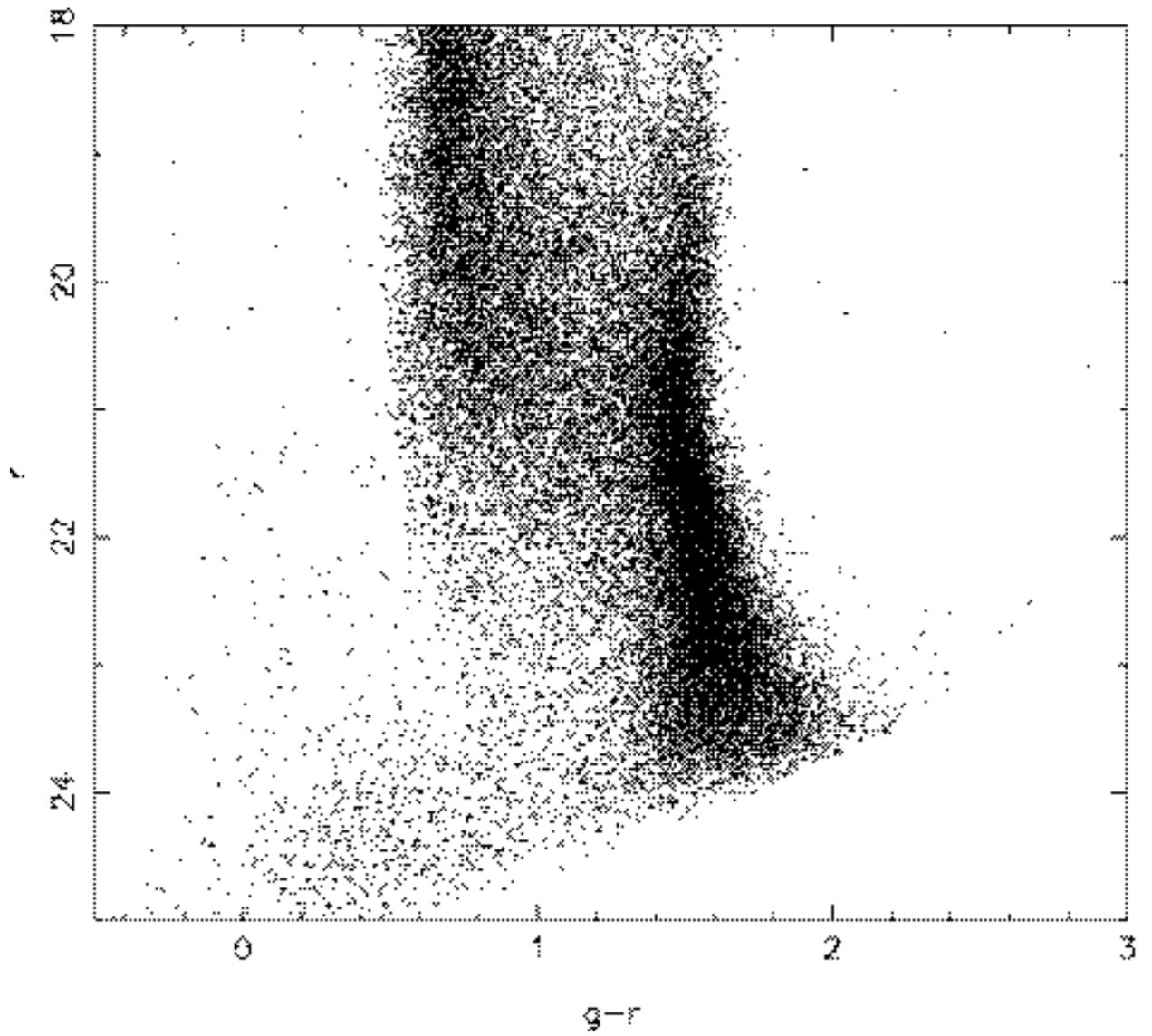,height=15cm,angle=-90}
\caption{A  $g,r$ colour-magnitude diagram  for objects  classified as
``stellar'' in the HVC region.
\label{Figure3}}            
\end{center}
\end{figure}

A  colour-magnitude diagram  for the  entire mosaic  region containing
images   classified  as   stellar  in   the  $r$-band   is   shown  in
figure~\ref{Figure3}.   The expected Galactic  components of  the thin
disk  K/M stars  are  clearly visible  as  the red  plume around  $g-r
\approx 1.5$, while the thick disk F/G stars make up the other obvious
component  at $g-r \approx  0.8$.  The  faint blue  stellar population
appearing  at the  bottom  left of  the  plot is  most  likely due  to
misclassified  blue compact  galaxies  at redshifts  $z \approx  0.5$.
This component is much  stronger in the equivalent non-stellar diagram
and in  both has  a fairly uniform  distribution over the  whole field
making it  unlikely to be related to  the HVC. Once again  there is no
compelling signature of any underlying dwarf galaxy component.

\section{Discussion}\label{discussion}
This paper has presented new  images of the extragalactic HI candidate
HIPASS \hvc, obtained with the WFI on the AAT. These are significantly
deeper than previous images,  reaching stellar magnitudes fainter than
$r = 24$ and $g = 25$.  Our search for any resolved associated stellar
component  was  negative.  Likewise,  there  is  no  evidence for  any
underlying  unresolved diffuse components  associated with  the Cloud.
Since  the  photon noise/  instrumental  signature surface  brightness
limits for detection  of unresolved structure are around  $r = 28$ and
$g = 29$ magnitudes arcsec$^{-2}$ respectively, we can place limits on
any  possible optical  counterpart.  In  practice, the  visibly patchy
Galactic  foreground nebulosity  becomes  important $\sim$1  magnitude
brighter than these limits; fortunately though, this patchiness avoids
the region where the HI contours are prominent.

Typical Local Group  dwarf galaxies have scale sizes  of order $\sim$1
kpc which, at a distance of  3\,Mpc, would be equivalent to an angular
scale  of  $\sim$1  arcmin.   We  would  expect  to  see  the  optical
counterpart of such  objects, if they exist, over  a few scale lengths
(cf. HI contours in  figure~\ref{Figure1}).  Considering a 1 arcminute
region and  our surface brightness  limits, the null detection  of any
diffuse optical structure leads to an absolute magnitude limit of $M_g
\approx -7$ for this  object, if it were at a distance  of 3 Mpc. This
is  comparable to the  limit derived  by Kilborn  et al.   (2000) from
photographic  plates  of  $M_B\approx-9$,  illustrating that  we  have
pushed  $\sim2$  magnitudes   further  down  the  luminosity  function
compared to this  earlier, shallower study.  This limit  is already at
the  lower  end of  the  range for  Local  Group  galaxies, making  it
unlikely than an optical component exists. Similarly, this lower limit
pushes the  inferred HI mass  to light ratio  to $>150$, at  least two
orders of magnitude encountered  in Local Group dwarf galaxies. Hence,
if \hvc\ is a Local Group  member, it represents a very different type
of object to what is currently known.

Such  conclusions make it  tempting to  support one  obvious alternate
scenario for \hvc\ -- that  it is a Galactic halo High-Velocity Cloud;
a recent search of POSS plates  for stellar content in a sample of 250
northern  HVCs yielded  no detections  (Simon \&  Blitz  2002).  While
there are a  few HVCs in the HIPASS catalog  (Putman et~al. 2002) with
Galactic  Standard of  Rest  velocities comparable  to  that of  \hvc,
should  \hvc\ be  a  halo HVC,  it  would likewise  necessarily be  an
extreme one.

It is  of further interest to  note that \hvc,  the earlier discovered
J1616$-$55  (Staveley-Smith  et~al 1998),  and  two additional  clouds
discovered  by  HIPASS  (Koribalski   et~al.   2002),  {\it  all}  lie
superimposed upon  the Supergalactic Plane (SGP).   An {\it extremely}
deep survey of  this region of the Supergalactic  Plane, searching for
any putative  connection between the clouds, is  currently underway at
Parkes (Koribalski, Gibson, et~al).

The observations  were processed using  a new pipeline  data reduction
facility at the AAO. This facility is based on the reduction procedure
developed for  data obtained with the  Wide Field Camera  on the Isaac
Newton  Telescope on  La  Palma.  Users  of  WFI at  the  AAT who  are
interested in  pipeline processing  their data should  contact Geraint
Lewis.

\section*{Acknowledgments}
M. Irwin thanks the AAO for their hospitality during a visit when much
of this work was finalised.


\section*{References}






\reference Blitz, L., Spergel, D. N., Teuben, P. J., Hartmann, D. {\&} 
Burton, W. B. 1999, ApJ, 514, 818

\reference Braun R., Burton W.~B., 1999, A\&A,  341, 437.

\reference Irwin, M.~\& Lewis, J.\ 2001, New Astronomy Review, 45, 105

\reference Kilborn, V.~A., et al.\ 2000, AJ, 120, 1342

\reference Klypin, A., Kravtsov, A.V., Valenzuela, O. \& Prada, F.
1999, ApJ, 522, 82

\reference Koribalski, B., et~al. 2002, in preparation

\reference Putman, M.E., et~al. 2002, AJ, in press

\reference Schlegel D.~J., Finkbeiner D.~P., Davis M., 1998, ApJ,  500, 525

\reference Sembach, K.~R., Gibson, B.~K., Fenner, Y. \& Putman, M.~E. 2002,
ApJ, in press

\reference Simon, J.~D. \& Blitz, L., 2002, ApJ, in press

\reference Staveley-Smith, L., et~al. 1998, AJ, 116, 2717

\reference Tinney, C.~G. 2001, AAO Newsletter, 96, 14

\end{document}